\documentclass[11pt,a4paper]{article}
\usepackage{jheppub}
\usepackage{epsfig}
\usepackage{amsfonts}
\usepackage{amscd}
\usepackage{latexsym}
\usepackage{amsmath,amssymb}
\usepackage{verbatim}
\usepackage{setspace}  
\usepackage{hyperref}
\usepackage{slashed}
\usepackage{float}
\usepackage{physics}
\usepackage{marvosym}
\usepackage[percent]{overpic}
\usepackage{stmaryrd}

\numberwithin{equation}{section}
\def\p{\partial}

\def\ca{{\mathcal A}}

\def\<{\langle }
\def\>{\rangle}
\def\eps{\varepsilon}

\def\zb{{\bar z}}
\def\oh{{\mathcal O}}
\def\real{{\mathbb R}}
\newcommand{\res}[1]{\underset{#1}{\rm Res}}

\newcommand{\ba}{\begin{aligned}}
\newcommand{\ea}{\end{aligned}}
\def\beq{\be\begin{array}{c}}
\def\eeq{\end{array}\ee} 
\def\be{\begin{equation}}
\def\ee{\end{equation}}
\def\bea{\begin{eqnarray}}
\def\eea{\end{eqnarray}}

\title{Celestial Locality and the Jacobi Identity}
 
\author{Adam Ball}

\affiliation{Department of Physics, Brown University,\\
Providence, RI 02912, USA\\
\rm{adam\_ball@brown.edu}}

\abstract{We show the equivalence of several different tests of the Jacobi identity for celestial currents at tree level, in particular finding a simple, practical condition on hard momentum space 4-point amplitudes in any EFT. Along the way we clarify the role of the order of soft and collinear limits in obstructing the Jacobi identity for soft insertions and we argue that, despite their current-algebra-like properties, soft insertions as formulated in this paper cannot be interpreted as local operators in celestial conformal field theory.}

\begin{document} 
\maketitle
\flushbottom

\section{Introduction} \label{sec:intro}

Taking inspiration from the holographic duality of quantum gravity in anti-de Sitter space, celestial conformal field theory (CCFT) aims to characterize 4D asymptotically flat quantum gravity by recasting its amplitudes as 2D correlation functions on the celestial sphere \cite{Pasterski:2016qvg}. For recent reviews, see \cite{Raclariu:2021zjz, Pasterski:2021rjz}. The program of celestial holography has seen remarkable progress in recent years, incorporating, among other things, universal aspects of scattering such as soft and collinear limits, which are respectively dual to soft currents and the celestial OPE \cite{Donnay:2018neh, Fan:2019emx, Pate:2019mfs, Adamo:2019ipt, Puhm:2019zbl, Pate:2019lpp, Himwich:2021dau}. The all-orders-soft current algebras for pure Yang-Mills theory and Einstein gravity were first displayed in \cite{Guevara:2021abz}, and it was soon realized in \cite{Strominger:2021lvk} that the modes of 2D light transforms of soft gravitons form a $w_{1+\infty}$-wedge current algebra. This algebra is unchanged by quantum effects in self-dual gravity \cite{Ball:2021tmb}. Despite these first few examples forming consistent algebras, it was shown in \cite{Mago:2021wje, Ren:2022sws} that the ostensible soft currents in many non-minimally coupled EFTs fail to satisfy the Jacobi identity. Other works involving the celestial Jacobi identity include \cite{Costello:2022wso, Costello:2022upu, Costello:2022jpg, Bittleston:2022jeq}. In this paper we continue the study of the Jacobi identity of celestial currents at tree level, establishing the equivalence of several conditions for it and clarifying the subtleties regarding the order of limits in the soft case. One of our main takeaways is that soft insertions as formulated below cannot be interpreted as local operators in CCFT,\footnote{There is a loophole in our argument if one allows only integer conformal dimensions.} despite the resemblance to a current algebra of their action on hard massless insertions. It may be that the appropriate local objects in CCFT are shadows of soft insertions, as in \cite{Kapec:2016jld, Kapec:2017gsg, Kapec:2021eug, Kapec:2022axw, Kapec:2022hih}.

The outline of the paper is as follows. In section \ref{sec:conv} we establish conventions. In section \ref{sec:hard} we briefly review the recent use of a double residue condition on hard momentum space amplitudes to test the Jacobi identity of celestial currents, and then we show using rather elementary tools that the double residue condition on hard momentum space amplitudes is fully equivalent to a simple condition on massless 4-point amplitudes. In section \ref{sec:ssoft} we review some properties of single soft insertions such as the locality of their poles and the nonlocality of their relationship with collinear limits. In section \ref{sec:msoft} we discuss some properties of multiple soft insertions, emphasizing the non-commutativity (and therefore nonlocality) of soft limits. Finally in section \ref{sec:jsoft} we discuss how to make the Jacobi identity for celestial soft currents well-defined, and we show that some reasonable definitions are equivalent to the aforementioned condition on hard amplitudes. We conclude in section \ref{sec:disc}.

\section{Conventions} \label{sec:conv}

There are several candidates in the celestial holography literature for the set of allowed values of the conformal dimension $\Delta$ \cite{Pasterski:2017kqt, Donnay:2020guq, Atanasov:2021oyu}. In this paper we are not primarily concerned with the question of completeness, so we allow $\Delta$ to be an arbitrary complex number. Mellin transforms encounter poles at integer values of $\Delta$, so when we write a generic $\Delta$ we intend it to be non-integer. There are many interesting proposals, including those in \cite{Pasterski:2017kqt, Sharma:2021gcz, Fan:2021pbp, Fan:2022vbz}, for alternate bases that involve integral transforms on the particles' angles, but in this paper we study only plane waves, conformal primary wavefunctions, and their soft limits. We refer to these objects as insertions in (celestial) amplitudes. We do not consider form factors nor amplitudes in nontrivial backgrounds.

We parametrize our massless momenta as
\be \label{eq:massless} p^\mu = \epsilon \, \omega \, q^\mu \ee
where $\epsilon = \pm 1$ determines whether the momentum is future- or past-directed and
\be q^\mu = (1+z\zb, z+\zb, -i(z-\zb), 1-z\zb). \ee
We parametrize a momentum with mass $m$ as
\be \label{eq:massive} p^\mu = \epsilon \, m \left( \frac{1+y^2+z\zb}{2y}, \frac{z+\zb}{2y}, \frac{-i(z-\zb)}{2y}, \frac{1-y^2-z\zb}{2y} \right). \ee
We use the all-outgoing convention for amplitudes, where crossing symmetry is used to trade any incoming particles for outgoing particles with past-directed momenta. Importantly, we treat $z, \zb$ as independent variables. This paper is only concerned with tree-level amplitudes, so we remain agnostic about whether our momenta are Lorentzian or Kleinian \cite{Atanasov:2021oyu}. To restrict to real Kleinian momenta, choose $z, \zb \in \real$ and Wick rotate the third component. Our choice of polarization vectors can be written simply in terms of $q^\mu$:
\be \ba \eps_+^\mu & \equiv \frac{1}{\sqrt{2}} \p_z q^\mu = \frac{1}{\sqrt{2}} (\zb, 1, -i, -\zb), \\
\eps_-^\mu & \equiv \frac{1}{\sqrt{2}} \p_\zb q^\mu = \frac{1}{\sqrt{2}} (z, 1, i, -z). \ea \ee
They obey the usual relations
\be 0 = \eps_\pm \cdot q = \eps_+ \cdot \eps_+ = \eps_- \cdot \eps_- \ee
and
\be \eps_+ \cdot \eps_- = 1. \ee

\section{Jacobi for hard insertions} \label{sec:hard}

A striking feature of the holomorphic celestial OPE \cite{Fan:2019emx, Pate:2019lpp} is that it always comes with a factor of $1/z$ \cite{Himwich:2021dau}, as opposed to some non-integer power of $z$, no matter the weights of the operators involved.\footnote{A simple example of an OPE lacking this property is that between two vertex operators in the free boson CFT.} In any CFT we are always free to analytically continue $z$ and $\zb$ separately, but in general there is no guarantee that the resulting $z$ dependence will be single-valued. The celestial OPE guarantees that in CCFT it \textit{is} single-valued, at least at leading order near massless insertions. Consequently the holomorphic OPEs of generic massless CCFT operators $\oh_{\Delta_i}(z_i, \zb_i)$ na\"ively resemble a holomorphic current algebra, whether or not the $\oh_{\Delta_i}(z_i, \zb_i)$ are soft. In light of this fact, one can ask whether these objects satisfy the current algebra version of the Jacobi identity, i.e. whether
\be \ba 0 & \stackrel{?}{=} \oint_{|z_{23}| = \eps} \frac{dz_2}{2\pi i} \oh_{\Delta_3}(z_3, \zb_3) \oint_{|z_{12}| = \eps} \frac{dz_1}{2\pi i} \oh_{\Delta_1}(z_1, \zb_1) \oh_{\Delta_2}(z_2, \zb_2) \\
& \qquad - \oint_{|z_{13}|=\eps} \frac{dz_1}{2\pi i} \oh_{\Delta_1}(z_1, \zb_1) \oint_{|z_{23}|=\eps} \frac{dz_2}{2\pi i} \oh_{\Delta_2}(z_2, \zb_2) \oh_{\Delta_3}(z_3, \zb_3) \\
& \qquad \qquad + \oint_{|z_{23}|=\eps} \frac{dz_2}{2\pi i} \oh_{\Delta_2}(z_2, \zb_2) \oint_{|z_{13}|=\eps} \frac{dz_1}{2\pi i} \oh_{\Delta_3}(z_3, \zb_3) \oh_{\Delta_1}(z_1, \zb_1) \ea \ee
where $z_{ij} = z_i - z_j$. This can be rewritten compactly as a ``double residue condition":
\be \label{eq:dblcel} 0 \stackrel{?}{=} \left( \res{z_2\shortrightarrow z_3} \, \res{z_1\shortrightarrow z_2} - \res{z_1\shortrightarrow z_3} \, \res{z_2\shortrightarrow z_3} + \res{z_2\shortrightarrow z_3} \, \res{z_1\shortrightarrow z_3} \right) \oh_{\Delta_1}(z_1, \zb_1) \oh_{\Delta_2}(z_2, \zb_2) \oh_{\Delta_3}(z_3, \zb_3). \ee
In both equations it is to be understood that the full celestial amplitude contains arbitrary other insertions away from $z_1, z_2, z_3$. Genuine currents in a local CFT are guaranteed to satisfy this condition by contour pulling, but the holomorphy of our celestial ``hard currents" is only known at leading order near massless insertions. Consequently the condition must be checked directly. Converting the collinear limit to the celestial OPE, as in \cite{Pate:2019lpp}, relies on the fact that taking a residue on $z_i\to z_j$ commutes with Mellin transforming on $\omega_i, \omega_j$.\footnote{See appendix \ref{app:cons} for a discussion of how to take residues of unstripped amplitudes.} This allows us to convert the double residue condition on celestial amplitudes to the same one on momentum space amplitudes:
\be \label{eq:dblmom} 0 \stackrel{?}{=} \left( \res{z_2\shortrightarrow z_3} \, \res{z_1\shortrightarrow z_2} - \res{z_1\shortrightarrow z_3} \, \res{z_2\shortrightarrow z_3} + \res{z_2\shortrightarrow z_3} \, \res{z_1\shortrightarrow z_3} \right)  \oh_1(\omega_1, z_1, \zb_1) \oh_2(\omega_2, z_2, \zb_2) \oh_3(\omega_3, z_3, \zb_3). \ee
This is the condition studied in \cite{Ren:2022sws}. They tested it for a large family of EFTs and found that it generically fails. We find that this failure can be traced to three-particle factorization channels and their associated nonlocal poles in $z_i$ in momentum space.\footnote{It is unclear what happens to a nonlocal pole in $z_i$ in momentum space upon transforming to Mellin space. Mellin transforming first on $\omega_i$ presumably gives some nonlocal branch point in $z_i$, but after all $\omega_j$ have been Mellin transformed it is unclear what kind of singularities in $z_i$ to expect.} By ``nonlocal" we mean that the poles in $z_i$ are not located at $z_j$ for any other insertion $\oh_j$. These stand in contrast with collinear poles, which are local.

\begin{figure}
  \centering
  \begin{minipage}[b]{0.31\textwidth}
    \centering
        \includegraphics[width=\textwidth]{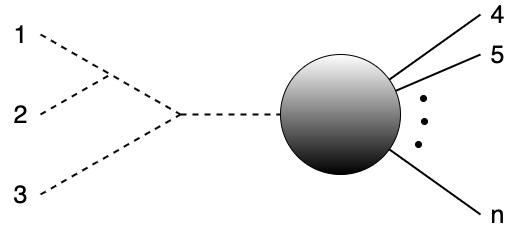}
  \end{minipage}
  \hfill
  \begin{minipage}[b]{0.31\textwidth}
  \centering
        \includegraphics[width=\textwidth]{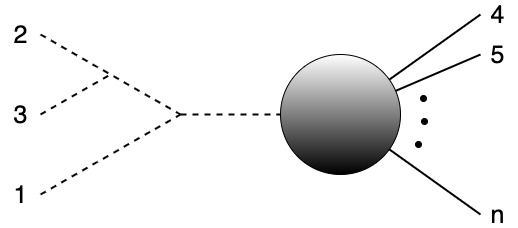}
  \end{minipage}
  \hfill
  \begin{minipage}[b]{0.31\textwidth}
  \centering
        \includegraphics[width=\textwidth]{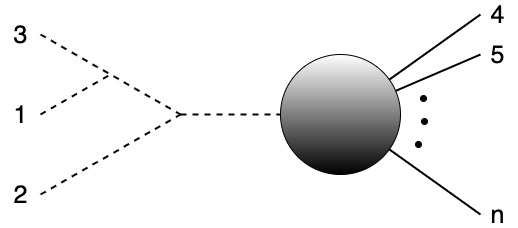}
  \end{minipage} 
  \caption{\label{fig:feyndias} Feynman diagrams that can contribute to the double residue condition on particles $1, 2, 3$. Dotted lines indicate massless bosons, solid lines indicate any particle in the theory, and the shaded spheres indicate any tree-level process.}
\end{figure}

To be concrete, let $\alpha_1, \alpha_2, \alpha_3$ label massless bosons of any helicity in any local, unitary, Lorentz-invariant EFT, and consider the tree-level $n\ge 6$-point momentum space amplitude $\ca_{\alpha_1\alpha_2\alpha_3\dots}$. View it as a rational function of $z_1, z_2, z_3$.\footnote{\label{ft:momelim} To make this well-defined choose two momenta $p_i, p_j$ with $i, j \ge 4$. If $p_i$ is massless use momentum conservation to eliminate $\omega_i$ and $\zb_i$, and if it is massive eliminate $y_i$ and $\zb_i$. Do the same for $p_j$. See appendix \ref{app:cons} for a more detailed discussion.} Generically the only poles near $z_1=z_2=z_3$ are the three $z_{12}, z_{13}, z_{23}$ poles from collinear factorization, and the pole from three-particle factorization where $(p_1+p_2+p_3)^2 = 0$. Only Feynman diagrams containing two such poles can contribute to the double residue condition. The candidates are shown in figure \ref{fig:feyndias}. They all factorize on $(p_1+p_2+p_3)^2=0$, so they all contain a factor of (a polynomial times) the following propagator,
\be \frac{1}{(p_1+p_2+p_3)^2} = \frac{-1/4}{\epsilon_1 \epsilon_2 \omega_1\omega_2z_{12}\zb_{12} + \epsilon_1 \epsilon_3 \omega_1\omega_3z_{13}\zb_{13} + \epsilon_2 \epsilon_3 \omega_2\omega_3z_{23}\zb_{23}}. \ee
This factor provides a pole at some $z_1=z_*$, but it is nonlocal in the sense that $z_* \ne z_i$ for any other $z_i$. This provides an obstruction to the contour pulling argument and can cause the double residue condition to fail. Before moving on let us briefly establish some notation. Thinking in terms of Feynman diagrams, write our amplitude as
\be \ca_{\alpha_1\alpha_2\alpha_3\dots} = \delta^{(4)}\left( \sum_{i=1}^n p_i \right) \left( \sum_{\hat\alpha_I} A^{\hat\alpha_I}_L \frac{1}{(p_1+p_2+p_3)^2} A^{\hat\alpha_I}_R \right) + \dots \ee
where the sum is over massless particles contributing to the $p_1+p_2+p_3$ factorization channel, $A^{\hat\alpha_I}_L$ collects the Feynman diagram factors to the left of the $\hat\alpha_I$ propagator,\footnote{When $\hat\alpha_I$ has spin then $A_{L/R}^{\hat\alpha_I}$ has suppressed indices.} $A_R^{\hat\alpha_I}$ collects the Feynman diagram factors to the right of the $\hat\alpha_I$ propagator, and ``$\dots$" indicates the contribution of diagrams not factorizing on $(p_1+p_2+p_3)^2=0$, which cannot contribute to the double residue condition. Note also that $A^{\hat\alpha_I}_R$ generically contains no poles in the neighborhood of $z_1=z_2=z_3$.

It was pointed out in \cite{Guevara:2022qnm} that the double residue condition \eqref{eq:dblmom} is equivalent to a second-order residue. There are always many equivalent ways to write a second-order residue in terms of first-order residues, and here one way turns out to be particularly illuminating. We find that
\be \label{eq:res} \res{z_2\shortrightarrow z_3} \res{z_1\shortrightarrow z_*} \ca_{\alpha_1\alpha_2\alpha_3\dots} = -\left( \res{z_2\shortrightarrow z_3} \, \res{z_1\shortrightarrow z_2} - \res{z_1\shortrightarrow z_3} \, \res{z_2\shortrightarrow z_3} + \res{z_2\shortrightarrow z_3} \, \res{z_1\shortrightarrow z_3} \right) \ca_{\alpha_1\alpha_2\alpha_3\dots}. \ee
This can be confirmed explicitly by plugging in a general expression for $A_L^{\hat\alpha_I}$. Recall that tree-level Feynman diagrams are rational functions of momentum components, with the denominators coming only from propagators. Thus for some polynomials $f^{\hat\alpha_I}_{ij}$ we can write
\be A^{\hat\alpha_I}_L = \frac{f^{\hat\alpha_I}_{12}}{(p_1+p_2)^2} + \frac{f^{\hat\alpha_I}_{13}}{(p_1+p_3)^2} + \frac{f^{\hat\alpha_I}_{23}}{(p_2+p_3)^2} + \dots \ee
where ``$\dots$" indicates massive exchange and contact terms, which have no poles in the neighborhood of $z_1=z_2=z_3$.

The left hand side of \eqref{eq:res} has a very clean interpretation. The first residue gives
\be \res{z_1\shortrightarrow z_*} \ca_{\alpha_1\alpha_2\alpha_3\dots} = \delta^{(4)}\left( \sum_{i=1}^n p_i \right) \sum_{\alpha_I} A_{\alpha_1\alpha_2\alpha_3\alpha_I} \frac{-1/4}{\epsilon_1 \omega_1 (\epsilon_2 \omega_2 \zb_{12} + \epsilon_3 \omega_3 \zb_{13})} A_{-\alpha_I\dots} \ee
where now $A^{\hat\alpha_I}_{L/R}$ have been replaced by bona fide (stripped) amplitudes and the sum is over helicities as well as particles. It is still true that $A_{-\alpha_I\dots}$ contains no poles in the neighborhood of $z_1=z_2=z_3$, so we can bring the second residue inside the sum as
\be \label{eq:insum} \res{z_2\shortrightarrow z_3} \, \res{z_1\shortrightarrow z_*} \ca_{\alpha_1\alpha_2\alpha_3\dots} = \delta^{(4)}\left( \sum_{i=1}^n p_i \right) \sum_{\alpha_I} \left( \res{z_2\shortrightarrow z_3} A_{\alpha_1\alpha_2\alpha_3\alpha_I} \right) \frac{-1/4}{\epsilon_1 \omega_1 (\epsilon_2 \omega_2 \zb_{12} + \epsilon_3 \omega_3 \zb_{13})} A_{-\alpha_I\dots}. \ee
Focus now on the residue term in parentheses. After taking the residue $z_1 \to z_*$ there is no more $z_1$ dependence, and we never introduced $p_I$ for $\alpha_I$, so the form of the 4-point amplitude $A_{\alpha_1\alpha_2\alpha_3\alpha_I}$ is as if we had used momentum conservation to eliminate $z_1$ and all three on-shell parameters of $p_I$. Then the only $z_{ij}$ that can show up in $A_{\alpha_1\alpha_2\alpha_3\alpha_I}$ is $z_{23}$, and the residue on $z_2\to z_3$ just grabs the coefficient of $1/z_{23}$ in the Laurent expansion in $z_{23}$. In spinor-helicity variables we would say that the only angle bracket allowed is $\langle 23 \rangle$, and the residue grabs from the $1/\langle 23 \rangle$ term. Applications of momentum conservation can make an amplitude unrecognizable, but they can never change the angle bracket weight, defined here as the number of angle bracket products in the numerator minus the number in the denominator.\footnote{In a case like $\frac{1}{m^2-\langle 23 \rangle [23]}$ we should expand as $\frac{1}{m^2} \sum_{n=0}^\infty \big(\langle 23 \rangle [23]/m^2\big)^n$ so that each term has definite angle bracket weight.} So no matter what form of $A_{\alpha_1\alpha_2\alpha_3\alpha_I}$ we are given, we can always read off the angle bracket weight $-1$ part as the obstruction to the double residue condition. We note that this is the lowest possible weight for a tree-level 4-point amplitude.

If the angle bracket weight $-1$ part of $A_{\alpha_1\alpha_2\alpha_3\alpha_I}$ vanishes for all $\alpha_I$, then \eqref{eq:insum} shows that the double residue condition will be satisfied for arbitrary ``\dots" in $\ca_{\alpha_1\alpha_2\alpha_3\dots}$. Conversely, suppose there exists an $\alpha_I$ such that the angle bracket weight $-1$ part of $A_{\alpha_1\alpha_2\alpha_3\alpha_I}$ is nonzero. Then there exists an $n\ge 6$-point amplitude failing the double residue condition, as follows. The idea is just to use CPT conjugates to get a sum over manifestly non-negative terms. Consider the 6-point amplitude $\ca_{\alpha_1\alpha_2\alpha_3\bar\alpha_1\bar\alpha_2\bar\alpha_3}$, where $\bar\alpha_1, \bar\alpha_2, \bar\alpha_3$ are the conjugate particles to $\alpha_1, \alpha_2, \alpha_3$. Use primes to denote the momenta of the $\bar\alpha_i$. We have
\be \res{z_2\shortrightarrow z_3} \, \res{z_1\shortrightarrow z_*} \ca_{\alpha_1\alpha_2\alpha_3\bar\alpha_1\bar\alpha_2\bar\alpha_3} = \frac{-\frac{1}{4} \delta^{(4)} \hspace{-1mm} \left( \sum_{i=1}^3 p_i + p'_i \right)}{\epsilon_1 \omega_1 (\epsilon_2 \omega_2 \zb_{12} + \epsilon_3 \omega_3 \zb_{13})} \sum_{\alpha_I} \left( \res{z_2\shortrightarrow z_3} A_{\alpha_1\alpha_2\alpha_3\alpha_I} \right) A_{\bar\alpha_I\bar\alpha_1\bar\alpha_2\bar\alpha_3}. \ee
We are free to take a further residue on $z'_2 \to z'_3$\footnote{See appendix \ref{app:cons} for a discussion of how to take residues of constrained functions.} and then set all $p'_i = -p_i$, which gives
\be \ba & \Big( \res{z'_2\shortrightarrow z'_3} \res{z_2\shortrightarrow z_3} \, \res{z_1\shortrightarrow z_*} \ca_{\alpha_1\alpha_2\alpha_3\bar\alpha_1\bar\alpha_2\bar\alpha_3} \Big) \Big|_{p'_i = -p_i} \\
& \hspace{20mm} = \frac{-\frac{1}{4} \delta^{(4)} \hspace{-1mm} \left( \sum_{i=1}^3 p_i + p'_i \right)}{\epsilon_1 \omega_1 (\epsilon_2 \omega_2 \zb_{12} + \epsilon_3 \omega_3 \zb_{13})} \sum_{\alpha_I} \Big( \res{z_2\shortrightarrow z_3} A_{\alpha_1\alpha_2\alpha_3\alpha_I} \Big) \Big( \res{z'_2\shortrightarrow z'_3} A_{\bar\alpha_I\bar\alpha_1\bar\alpha_2\bar\alpha_3} \Big) \Big|_{p'_i = -p_i} \\
& \hspace{20mm} = \frac{-\frac{1}{4} \delta^{(4)}( 0 )}{\epsilon_1 \omega_1 (\epsilon_2 \omega_2 \zb_{12} + \epsilon_3 \omega_3 \zb_{13})} \sum_{\alpha_I} \Big| \res{z_2\shortrightarrow z_3} A_{\alpha_1\alpha_2\alpha_3\alpha_I} \Big|^2. \ea \ee
Since we are assuming that at least one $\res{z_2\shortrightarrow z_3} A_{\alpha_1\alpha_2\alpha_3\alpha_I}$ is nonzero, this sum cannot vanish. This completes the proof of the following:\\

\noindent \textbf{Theorem} \quad The double residue condition \eqref{eq:dblmom} on massless bosons $\alpha_1, \alpha_2, \alpha_3$ fails if and only if there exists a massless boson $\alpha_I$ such that the angle bracket weight $-1$ part of the 4-point amplitude $A_{\alpha_1\alpha_2\alpha_3\alpha_I}$ is nonzero.\\

\noindent This strengthens and generalizes some of the observations in \cite{Ren:2022sws}. In particular the angle bracket weight $-1$ part of a 4-point amplitude is equivalent to its all-line shift constructible part. This condition also applies directly to the holographic chiral algebras of \cite{Monteiro:2022lwm}.

Sometimes it is more useful to work with momentum weight rather than angle bracket weight, where e.g. $p_i \cdot p_j$ has momentum weight $+2$ and $\eps_{\pm,i} \cdot p_j$ has momentum weight $+1$. Let $W_a$ denote angle bracket weight, $W_s$ denote square bracket weight, and $W_m$ denote momentum weight. They are related as $W_m = W_a + W_s$. Furthermore little group scaling implies $W_a - W_s = -\sum_i s_i$, where $s_i$ is the helicity of the $i$th external leg of our amplitude. Combining these gives
\be W_m = 2W_a + \sum_i s_i, \ee
so that angle bracket weight $W_a=-1$ corresponds to momentum weight
\be W_m = -2 + \sum_i s_i. \ee

\section{Single soft insertions} \label{sec:ssoft}

In this section we review some of the properties of single soft insertions. Let $\oh(\omega, z, \zb)$ denote a massless insertion in a momentum space amplitude where all other insertions are hard. It is a well-defined rational function of $\omega, z, \zb$ once we enforce momentum conservation as discussed in appendix \ref{app:cons}. In doing so, for convenience choose not to eliminate any of the $z_i$ of the other insertions. We can Laurent expand near $\omega=0$, giving
\be \oh(\omega, z, \zb) = \sum_{k=-\infty}^1 \frac{\oh^{(k)}(z, \zb)}{\omega^k}. \ee
The $\oh^{(k)}(z, \zb)$ are energetically soft insertions. For later convenience we will also define $S^{(k)}$ to grab the coefficient of $\omega^{-k}$ in the Laurent series around $\omega=0$ of whatever function it is acting on. Then by definition
\be S^{(k)} \oh(\omega, z, \zb) = \oh^{(k)}(z, \zb). \ee
As discussed in \cite{Guevara:2019ypd}, there is a theorem stating that the Laurent coefficient $\oh^{(k)}(z, \zb)$ is equal to the residue at $\Delta \to k$ of the Mellin transform of $\oh(\omega, z, \zb)$. That is, the energetically and conformally soft limits give the same objects, at least in this context of handling one insertion at a time. In particular, $\oh^{(k)}(z, \zb)$ is a conformal primary despite being constructed from momentum space. Note also that its $z$ dependence is manifestly single-valued around momentum space insertions. Sometimes we will abuse notation and let $S^{(k)}$ act on a function of $\Delta$ rather than $\omega$, in which case it should be understood that it corresponds to the residue at $\Delta \to k$. As a final preliminary comment, note that rational functions are equal to their Laurent expansions, so $\oh$ can be recovered from its soft modes $\oh^{(k)}$ and in this sense they form a complete set of insertions.

Since $\omega \oh(\omega, z, \zb)$ is analytic in $\omega$ near $\omega=0$, we can compute the Laurent coefficient $S^{(k)} \oh(\omega, z, \zb)$ by acting on $\oh(\omega, z, \zb)$ with the differential operator $\frac{1}{(1-k)!} \p_\omega^{1-k} \omega$ and then setting $\omega=0$. We know that $\omega \oh(\omega, z, \zb)$ is a rational function of $\omega, z, \zb$, so it can be written as a ratio of polynomials in $\omega, z, \zb$. Differentiating by $\omega$ cannot produce any new factors in the denominator; it can only increase the powers of factors that are already there. So the only possible $z$ poles (of any order) in $\oh^{(k)}(z, \zb)$ are those corresponding to $z$ poles in $\omega \oh(\omega, z, \zb)|_{\omega=0}$. But $z$ poles in $\omega\oh(\omega, z, \zb)$ only come from two-particle factorization, higher-particle factorization, and places where the coordinates degenerate. The latter two types of poles go away when $\omega=0$ since $p(\omega, z, \zb)$ simply drops out of the expressions. Therefore the only poles in $\oh^{(k)}(z, \zb)$ are those coming from two-particle factorization. In the massless case these are just the familiar collinear poles. In the massless-massive case they come from the massive propagator
\be \frac{1}{m^2 + \big( p(\omega, z, \zb) + p(y, z', \zb'; m) \big)^2} = \frac{1}{2p(\omega, z, \zb) \cdot p(y, z', \zb'; m)} = \frac{y}{-2m\epsilon \epsilon' \omega (y^2 + |z-z'|^2)}. \ee
We see this has a pole at $z = z' - \frac{y^2}{\zb-\zb'}$. Technically this is nonlocal in this paper's parlance, but its location does not depend on any of the other massless $z_i$, so it is irrelevant to computing the double residue condition on massless insertions. Now that we have deduced all the $z$ poles of the rational function $\oh^{(k)}(z, \zb)$, we can write it as a sum over poles plus a polynomial in $z$,\footnote{This expression is for a momentum space amplitude, but changing to a conformal primary basis for the hard insertions would leave much of it qualitatively unchanged. In particular there would still be a sum over massless poles, a sum over terms from massless-massive two-particle factorization, and a polynomial term collecting the rest of the $z$ dependence.}
\be \label{eq:softpolestruc} \oh^{(k)}(z, \zb) = \sum_i \frac{f_i^{(k)}(\zb)}{z-z_i} + \sum_j \frac{g_j^{(k)}(\zb)}{z - z_j + \frac{y_j^2}{\zb - \zb_j}} + P^{(k)}(z, \zb). \ee
Here $i$ ranges over the massless hard momenta, and $f_i^{(k)}(\zb)$ is determined by soft-collinear limits. Likewise $j$ ranges over the massive momenta, and $g_i^{(k)}(\zb)$ is determined by massless-massive two-particle factorization channels. Although $P^{(k)}(z, \zb)$ is polynomial in $z$, it is not necessarily polynomial in $\zb$. In the cases where universal soft theorems apply we will have $P^{(k)}(z, \zb) = 0$, but this is not the case in general. The point here is that we have substantial analytic control over the $z$ dependence of single soft insertions.

The soft and collinear limits $S^{(k)}$ and $\res{z\shortrightarrow z_i}$ (where particle $i$ is massless) commute since $\omega (z-z_i) \oh(\omega, z, \zb)$ is analytic in $\omega, z$ in a neighborhood of $\omega=0, z=z_i$, so the differential operator implementing $S^{(k)}$ commutes with setting $z=z_i$. This commutativity is implicit in \cite{Guevara:2021abz, Himwich:2021dau, Mago:2021wje, Ren:2022sws}. Note that the $\zb$ modes of $\oh^{(k)}(z, \zb)$ will be holomorphic with simple poles coming only from collinear limits (and from massless-massive two-particle factorization that we do not care about in this context). As argued in \cite{Guevara:2021abz}, these $\oh^{(k)}(z, \zb)$ are tantamount to symmetry-generating currents. However we will argue below that they cannot actually be interpreted as local operators in CCFT.

As an example, consider two positive-helicity outgoing gluons $\oh^{+,a}(\omega_1, z_1, \zb_1)$ and $\oh^{+,b}(\omega_2, z_2, \zb_2)$ in pure Yang-Mills. Assume the variables $\omega_1, z_1, \zb_1, \omega_2, z_2, \zb_2$ have not been eliminated. The collinear limit is
\be \oh^{+,a}(\omega_1, z_1, \zb_1) \oh^{+,b}(\omega_2, z_2, \zb_2) \sim \frac{-i{f^{ab}}_c}{z_{12}} \frac{\omega_1 + \omega_2}{\omega_1 \omega_2} \oh^{+,c} \big( \omega_1 + \omega_2, z_2, \frac{\omega_1 \zb_1 + \omega_2 \zb_2}{\omega_1 + \omega_2} \big). \ee
To compute the collinear limit with the soft insertion $[\oh^{+,a}]^{(0)}$ we expand in $\omega_1$ and grab the $O(\omega_1^0)$ piece, giving
\be [\oh^{+,a}]^{(0)}(z_1, \zb_1) \oh^{+,b}(\omega_2, z_2, \zb_2) \sim \frac{-i{f^{ab}}_c}{z_{12}} \left( \p_{\omega_2} + \frac{1}{\omega_2} + \frac{\zb_{12}}{\omega_2} \p_{\zb_2} \right) \oh^{+,c}(\omega_2, z_2, \zb_2). \ee
The term in parentheses is the subleading soft gluon factor \cite{Casali:2014xpa}. The dependence on $\zb_1$ is linear, and in general the $\zb_1$ dependence of the collinear limit of $[\oh^{+,a}]^{(k)}(z_1, \zb_1) \oh^{+,b}(\omega_2, z_2, \zb_2)$ will be polynomial of degree $1-k$. Consequently if we expand $[\oh^{+,a}]^{(k)}(z_1, \zb_1)$ in $\zb_1$ then the holomorphic coefficients of $\zb_1^m$ with $m<0$ and $m>1-k$ will not see this collinear pole. We sometimes say that such modes are ``outside the wedge". Very similar statements hold for soft-collinear limits in general EFTs. Finally we note that Mellin transforming on $\omega_2$ recovers the soft-hard celestial OPE \cite{Guevara:2021abz}, which takes the following closed form for general $k$,
\be [\oh^{+,a}]^{(k)}(z_1, \zb_1) \oh^{+,b}_{\Delta_2}(z_2, \zb_2) \sim \frac{-i{f^{ab}}_c}{z_{12}} \sum_{m=0}^{1-k} {2-k-\Delta_2-m \choose 1-\Delta_2} \frac{\zb_{12}^m}{m!} \p_{\zb_2}^m \oh^{+,c}_{\Delta_2+k-1}(z_2, \zb_2), \ee
where the binomial coefficient is defined as ${x \choose y} \equiv \frac{\Gamma(x+1)}{\Gamma(y+1) \Gamma(x-y+1)}$.

Despite the striking resemblance of the $z_1$ dependence of $\oh^{(k)}(z_1, \zb_1)$ to that of a holomorphic current in CFT, $\oh^{(k)}(z_1, \zb_1)$ cannot be interpreted as a local operator in CCFT. If it were local then the OPE of two operators far from it could not be affected by it, meaning that in pure Yang-Mills we would have
\be \ba \res{z_1\shortrightarrow z_2} & \left( \oh^{+,a_1}_{\Delta_1}(z_1, \zb_1) \oh^{+,a_2}_{\Delta_2}(z_2, \zb_2) [\oh^{+,a_3}]^{(k)}(z_3, \zb_3) \right) \stackrel{?}{=} \\
& -i{f^{a_1a_2}}_{b} \sum_{m=0}^\infty B(\Delta_1-1+m, \Delta_2-1) \frac{\zb_{12}^m}{m!} \p_{\zb_2}^m \oh^{+,b}_{\Delta_1+\Delta_2-1}(z_2, \zb_2) [\oh^{+,a_3}]^{(k)}(z_3, \zb_3). \ea \ee
But the left hand side is defined by taking the soft limit before the residue, while the right hand side is what we would get if we took the residue first. Equality would imply that the residue and the soft limit commute, which is not true even in pure Yang-Mills. If we consider a celestial MHV amplitude with color ordering $\dots \oh^+_{\Delta_1}(z_1, \zb_1) \oh^{+,(k)}(z_3, \zb_3) \oh^+_{\Delta_2}(z_2, \zb_2) \dots$ then taking ${\rm Res}_{z_1\shortrightarrow z_2}$ first gives zero whereas taking the soft limit first gives something nonzero. In general this happens because when a particle goes soft its three-particle factorization channels become two-particle factorization channels for the remaining hard particles. But the physical simplicity does not change the fact that it is inconsistent with $\oh^{(k)}$ being a local operator in CCFT.\footnote{If one allows only integer values of $\Delta$ then the soft limit is part of the definition of CCFT, as opposed to being a limit of an object within CCFT, and so the preceding argument may not apply.} We will say more about the nonlocality of soft insertions in the following sections.

\section{Multiple soft insertions} \label{sec:msoft}

In this section we discuss properties of amplitudes in which multiple insertions have been taken soft. Once we choose an order of limits for a momentum space amplitude, e.g.
\be S_3^{(k_3)} S_2^{(k_2)} S_1^{(k_1)} \oh_1(\omega_1, z_1, \zb_1) \oh_2(\omega_2, z_2, \zb_2) \oh_3(\omega_3, z_3, \zb_3), \ee
then many of the properties of single soft insertions carry through. In particular we have rational dependence on $z_1, z_2, z_3$ and poles only from two-particle factorization, although now they need not be simple poles. In general the soft limits do not commute, i.e.
\be S_1^{(k)} S_2^{(\ell)} \oh_1(\omega_1, z_1, \zb_1) \oh_2(\omega_2, z_2, \zb_2) \ne S_2^{(\ell)} S_1^{(k)} \oh_1(\omega_1, z_1, \zb_1) \oh_2(\omega_2, z_2, \zb_2). \ee
Once again this presents an obstacle to interpreting soft insertions as local operators in CCFT, as there is no sense of the ``order of insertion" of local operators \cite{Kapec:2022hih}. The commutators of some of the most leading soft limits for gluons and gravitons have been studied in detail in \cite{Lipstein:2015rxa, Klose:2015xoa, Anupam:2018vyu, Distler:2018rwu, Fotopoulos:2019vac, Fotopoulos:2020bqj, Campiglia:2021bap}. In particular, for positive-helicity gluons in pure Yang-Mills we have $[S_1^{(1)}, S_2^{(1)}] = 0$ and $[S_1^{(1)}, S_2^{(0)}] = 0$, with $[S_1^{(0)}, S_2^{(0)}]$ and $[S_1^{(1)}, S_2^{(-1)}]$ being nonzero. All further subleading commutators are nonzero as well. Opposite-helicity gluons in Yang-Mills fail to commute even at leading order, $[S_1^{(1)}, S_2^{(1)}] \ne 0$. This latter fact was appreciated in \cite{He:2015zea}, where it was recognized as an obstruction to defining the OPE of two opposite-helicity soft gluon currents. Recently some authors have found it convenient to use the simultaneous conformally soft limit  \cite{Mago:2021wje, Ren:2022sws}
\be \label{eq:simsoft} \lim_{\eps\to 0} \eps^2 \oh_{\Delta_1=k+\eps}(z_1, \zb_1) \oh_{\Delta_2=\ell+\eps}(z_2, \zb_2), \ee
which treats the two insertions symmetrically. More generally one could use arbitrary relative rates of softness,
\be \lim_{\eps\to 0} \eta_1\eta_2\eps^2 \oh_{k+\eta_1\eps}(z_1, \zb_1) \oh_{\ell+\eta_2\eps}(z_2, \zb_2). \ee
One can recover either consecutive limit by sending $\eta_2/\eta_1$ to zero or infinity, but for generic $\eta_1, \eta_2$ it is not clear how this limit is related to energetically soft limits. It is instructive to apply this limit to the celestial OPE in pure Yang-Mills. We will see that the non-commutativity is rather tame. For two positive-helicity outgoing gluons one finds
\be \ba & \lim_{\eps\to 0} \eta_1\eta_2\eps^2 \oh^{+,a}_{k+\eta_1\eps}(z_1, \zb_1) \oh^{+,b}_{\ell+\eta_2\eps}(z_2, \zb_2) \sim \\
& \frac{-i{f^{ab}}_c}{z_{12}} \left[ \sum_{m=0}^{1-k} {2 \hspace{-.7mm} - \hspace{-.7mm} k \hspace{-.7mm} - \hspace{-.7mm} \ell \hspace{-.7mm} - \hspace{-.7mm} m \choose 1 \hspace{-.7mm} - \hspace{-.7mm} \ell} + \frac{(-)^{1-\ell}}{1 \hspace{-.7mm} + \hspace{-.7mm} \eta_2/\eta_1} \sum_{m=3-k-\ell}^\infty {k \hspace{-.7mm} - \hspace{-.7mm} 2 \hspace{-.7mm} + \hspace{-.7mm} m \choose 1 \hspace{-.7mm} - \hspace{-.7mm} \ell} \right] \frac{\zb_{12}^m}{m!} \p_{\zb_2}^m [\oh^{+,c}]^{(k+\ell-1)}(z_2, \zb_2). \ea \ee
The $\eta_2/\eta_1$-dependent part is not necessarily zero, but it cannot have any poles at all in $z_2$ since all collinear poles would have polynomial $\zb_2$ dependence of degree $1-(k+\ell-1) = 2-k-\ell$, and this is always killed by $\p_{\zb_2}^m$ for $m \ge 3-k-\ell$. For two opposite-helicity outgoing gluons one finds
\be \ba & \lim_{\eps\to 0} \eta_1\eta_2\eps^2 \oh^{+,a}_{k+\eta_1\eps}(z_1, \zb_1) \oh^{-,b}_{\ell+\eta_2\eps}(z_2, \zb_2) \sim \\
& \frac{-i{f^{ab}}_c}{z_{12}} \left[ \sum_{m=0}^{1-k} {-k \hspace{-.7mm} - \hspace{-.7mm} \ell \hspace{-.7mm} - \hspace{-.7mm} m \choose -1 \hspace{-.7mm} - \hspace{-.7mm} \ell} + \frac{(-)^{-1-\ell}}{1 \hspace{-.7mm} + \hspace{-.7mm} \eta_2/\eta_1} \sum_{m=1-k-\ell}^\infty {k \hspace{-.7mm} - \hspace{-.7mm} 2 \hspace{-.7mm} + \hspace{-.7mm} m \choose -1 \hspace{-.7mm} - \hspace{-.7mm} \ell} \right] \frac{\zb_{12}^m}{m!} \p_{\zb_2}^m [\oh^{-,c}]^{(k+\ell-1)}(z_2, \zb_2). \ea \ee
Similar comments apply, with the $\eta_2/\eta_1$-dependent part having no poles in $z_2$.

\section{Jacobi for soft insertions} \label{sec:jsoft}

The soft double residue condition might na\"ively be written as
\be \label{eq:softcond} \left( \res{z_2\shortrightarrow z_3} \, \res{z_1\shortrightarrow z_2} - \res{z_1\shortrightarrow z_3} \, \res{z_2\shortrightarrow z_3} + \res{z_2\shortrightarrow z_3} \, \res{z_1\shortrightarrow z_3} \right) \oh_1^{(k_1)}(z_1, \zb_1) \oh_2^{(k_2)}(z_2, \zb_2) \oh_3^{(k_3)}(z_3, \zb_3), \ee
but this expression is ill-defined due to the non-commutativity of soft limits with each other and more importantly with collinear limits. We discuss three ways to make sense of this expression. First we address the one that has already been discussed in the literature, which is to choose some definition for the soft-soft OPE of two insertions and apply it in succession to define \eqref{eq:softcond}. In \cite{Mago:2021wje} the soft-soft OPE was defined using the simultaneous soft limit \eqref{eq:simsoft}, which treats the two insertions symmetrically. But with this definition, even positive-helicity gluons in pure Yang-Mills fail the double residue condition.\footnote{The offending terms are only beyond the wedge.} In \cite{Guevara:2021abz} the soft-soft OPE of $\oh_1\oh_2$ was defined by starting from the hard OPE centered on $z_2$, taking $\Delta_1$ and $\Delta_2$ soft, and discarding the beyond-wedge part by hand, which is equivalent to taking $\Delta_1$ soft first. This treats the two insertions asymmetrically (even after resummation), but it does lead to positive-helicity gluons in pure Yang-Mills satisfying \eqref{eq:softcond}.

Another way to define \eqref{eq:softcond} is by choosing some order of soft limits and taking them before the residues, e.g.
\be \label{eq:softfirst} \left( \res{z_2\shortrightarrow z_3} \, \res{z_1\shortrightarrow z_2} - \res{z_1\shortrightarrow z_3} \, \res{z_2\shortrightarrow z_3} + \res{z_2\shortrightarrow z_3} \, \res{z_1\shortrightarrow z_3} \right) S_3^{(k_3)} S_2^{(k_2)} S_1^{(k_1)} \oh(\omega_1, z_1, \zb_1) \oh(\omega_2, z_2, \zb_2) \oh(\omega_3, z_3, \zb_3). \ee
As discussed above, the $z_1, z_2, z_3$ dependence is rational with only local poles near $z_1=z_2=z_3$. (The massless-massive two-particle factorization poles are absent in the neighborhoods of generic points on the $z_1=z_2=z_3$ submanifold of parameter space.) Then by contour pulling, the double residue condition must be satisfied, even in EFTs with arbitrary non-minimal couplings. One naturally wonders what the corresponding modification to the celestial soft current algebra is in such theories, but this question appears ill-posed. If we tried to read off commutators, then the commutators of modes of $\oh_2$ and $\oh_3$ would depend on the properties of $\oh_1$. This is a manifestation of the nonlocality of soft insertions discussed above.

Finally we can define \eqref{eq:softcond} by taking soft limits after the residues. Note then that if the hard double residue condition is satisfied, we must get zero. Consider arbitrary relative rates of conformal softness for the three insertions,
\be\lim_{\eps\to 0} \eta_1\eta_2\eta_3 \eps^3 \bigg( \res{z_2\shortrightarrow z_3} \res{z_1\shortrightarrow z_2} \hspace{-.7mm} - \hspace{-.7mm} \res{z_1\shortrightarrow z_3} \res{z_2\shortrightarrow z_3} \hspace{-.7mm} + \hspace{-.7mm} \res{z_2\shortrightarrow z_3} \res{z_1\shortrightarrow z_3} \bigg) \oh_{k_1+\eta_1\eps}(z_1, \zb_1) \oh_{k_2+\eta_2\eps}(z_2, \zb_2) \oh_{k_3+\eta_3\eps}(z_3, \zb_3). \ee
Each residue splits into ``wedge" and ``beyond-wedge" parts identical to the simultaneous soft OPE up to $\eta_i$ dependence. We leave this $\eta_i$ dependence explicit in the following but otherwise adopt a very compact notation:
\be \ba \lim_{\eps\to 0} \eta_1\eta_2\eta_3 \eps^3 \res{z_2\shortrightarrow z_3} \res{z_1\shortrightarrow z_2} \oh_1\oh_2\oh_3 & = \left( W_{1,2} + \frac{\eta_1}{\eta_1+\eta_2} B_{1,2} \right) \left( W_{12,3} + \frac{\eta_1+\eta_2}{\eta_1+\eta_2+\eta_3} B_{12,3} \right), \\
\lim_{\eps\shortrightarrow 0} \eta_1\eta_2\eta_3 \eps^3 \res{z_1\shortrightarrow z_3} \res{z_2\shortrightarrow z_3} \oh_1\oh_2\oh_3 & = \left( W_{2,3} + \frac{\eta_2}{\eta_2+\eta_3} B_{2,3} \right) \left( W_{1,23} + \frac{\eta_1}{\eta_1+\eta_2+\eta_3} B_{1,23} \right), \\
\lim_{\eps\shortrightarrow 0} \eta_1\eta_2\eta_3 \eps^3 \res{z_2\shortrightarrow z_3} \res{z_1\shortrightarrow z_3} \oh_1\oh_2\oh_3 & = \left( W_{1,3} + \frac{\eta_1}{\eta_1+\eta_3} B_{1,3} \right) \left( W_{2,13} + \frac{\eta_2}{\eta_1+\eta_2+\eta_3} B_{2,13} \right). \ea \ee
Note that there is no choice of $\eta_1, \eta_2, \eta_3$ such that all of the $\eta_i$ ratios are equal to one half, which would correspond to using the simultaneous soft OPE. Terms in the double residue condition with linearly independent $\eta_i$ dependence must vanish separately. In this way one finds six independent equations satisfied by the $W$'s and $B$'s. These same six equations are already implied by the six possible consecutive soft limits on $\Delta_1, \Delta_2, \Delta_3$, so it turns out that the more general limit considered here did not give any extra information. The consecutive soft limit $S^{(k_3)}_3 S_2^{(k_2)} S_1^{(k_1)}$ is particularly interesting because it involves only the wedge terms,\footnote{There are many ways to rewrite the double residue condition by swapping $\res{z_i\shortrightarrow z_j}$ for $\res{z_j\shortrightarrow z_i}$, and different choices will lead to different consecutive soft limits being associated with the wedge terms.}
\be \ba \label{eq:softwedge} & W_{1,2} W_{12,3} - W_{2,3} W_{1,23} + W_{1,3} W_{2,13} = \\
& S_3^{(k_3)} S_2^{(k_2)} S_1^{(k_1)} \left( \res{z_2\shortrightarrow z_3} \, \res{z_1\shortrightarrow z_2} - \res{z_1\shortrightarrow z_3} \, \res{z_2\shortrightarrow z_3} + \res{z_2\shortrightarrow z_3} \, \res{z_1\shortrightarrow z_3} \right) \oh_{\Delta_1}(z_1, \zb_1) \oh_{\Delta_2}(z_2, \zb_2) \oh_{\Delta_3}(z_3, \zb_3). \ea \ee
Since the the soft limits (once they are moved inside the Mellin transforms) are just grabbing Laurent coefficients, the only way for $S^{(k_3)}_3 S_2^{(k_2)} S_1^{(k_1)}$ to vanish for all $k_1, k_2, k_3$ is if it acts on zero. Therefore the hard double residue condition is satisfied if and only if the wedge part of the soft double residue condition is satisfied. This is consistent with the results of \cite{Mago:2021wje, Ren:2022sws}, which found for a large family of EFTs that the constraints of the hard double residue condition are the same as those of one formulation of the soft double residue condition. Comparing with \eqref{eq:softfirst}, which vanishes, shows that any failure of \eqref{eq:softwedge} can be traced to non-commutativity of soft and collinear limits.

It turns out that $B_{1,2} W_{12,3}$ always vanishes whether or not the hard double residue condition is satisfied, simply because the powers of $\p_{\zb_2}$ in $B_{1,2}$ annihilate the $\zb_2$ dependence of $W_{12,3}$. This means that no matter what $\eta_1, \eta_2$ are, if we take $\eta_3\to\infty$ first (i.e. take $\oh_3$ soft last) then we will get the left hand side of \eqref{eq:softwedge}. This is still somewhat unsatisfactory because it treats $\oh_3$ differently from $\oh_1$ and $\oh_2$, despite all three nominally being the same type of soft object. We advocate for simply leaving $\oh_3$ hard, in which case the double residue condition amounts to equivariance of the action of $\oh_1^{(k_1)}, \oh_2^{(k_2)}$ on $\oh_3$. That is, if $X_1$ is a soft mode of $\oh_1$ and $X_2$ is a soft mode of $\oh_2$, then the double residue condition informs us about whether
\be X_1 \cdot (X_2 \cdot \oh_3) - X_2 \cdot (X_1 \cdot \oh_3) \stackrel{?}{=} [X_1, X_2] \cdot \oh_3, \ee
where $X_i \cdot \oh_3$ is defined through the soft-hard OPE. The commutator $[X_1, X_2]$ is computed with the soft-soft OPE, and any ambiguity from the order of soft limits drops out of $[X_1, X_2] \cdot \oh_3$. This is similar to the approach taken in \cite{Himwich:2021dau} in the context of $w_{1+\infty}$ generators. The upshot is that, when the double residue condition is satisfied, the consecutive action of soft insertions on a hard insertion is consistent with that of a current algebra, despite the fact that soft insertions as formulated in this paper are not truly local operators in CCFT.

\section{Discussion} \label{sec:disc}

In this paper we studied the Jacobi identity for holomorphic celestial currents at tree level in the form of the double residue condition. We showed that the question of its satisfaction has the same answer for hard insertions in Mellin space, hard insertions in momentum space, and (suitably defined) soft insertions. We further established its equivalence with a simple, practical condition on massless 4-point momentum space amplitudes: the vanishing of the angle bracket weight $-1$ part. This condition facilitates the application of known amplitudes results to questions in celestial holography.

We also highlighted the important role of the order of limits in obstructing the Jacobi identity for soft insertions. We discussed three different approaches to defining the soft double residue condition, and advocated for simply leaving one insertion hard and viewing the condition as a statement about equivariance of the action of the soft insertions on this hard insertion. This involves only terms within the wedge. When this double residue condition is satisfied, the action on the hard insertion is consistent with 2D locality. But other properties of the soft insertions, related to non-commutativity of limits, are simply incompatible with 2D locality, even for the relatively simple case of positive-helicity gluons in pure Yang-Mills.\footnote{As discussed above, there is a loophole if one only allows integer values of $\Delta$.} This observation is supported by the results of \cite{Kapec:2016jld, Kapec:2017gsg, Kapec:2021eug, Kapec:2022axw, Kapec:2022hih}, which suggest that it is actually the 2D shadows of soft insertions that play the role of local currents. It is argued in \cite{Kapec:2022hih} that, at least for the leading soft theorems, soft gluons and gravitons should be thought of as integrated operators whose insertion deforms the CCFT in conformal perturbation theory. Then the commutators of (leading) soft limits correspond to curvature on the conformal manifold of CCFTs. We leave the study of multiple insertions of soft shadows to the future, but we note that \cite{Fan:2021pbp, Hu:2022syq, De:2022gjn, Chang:2022jut} constitute preliminary work in this direction.

\acknowledgments

We thank Akshay Yelleshpur Srikant for invaluable discussions during the development of this work, and Andrew Strominger for comments on the draft. We also thank Eduardo Casali, Alfredo Guevara, Elizabeth Himwich, Lecheng Ren, and Anastasia Volovich for useful conversations. We gratefully acknowledge support from Simons Investigator Award \#376208 of A. Volovich.

\appendix

\section{Enforcing momentum conservation} \label{app:cons}

We denote an unstripped momentum space amplitude with $m$ massless legs and $n-m$ massive legs as
\be \ca_n(\omega_1, z_1, \zb_1, \dots, \omega_m, z_m, \zb_m, y_{m+1}, z_{m+1}, \zb_{m+1}, \dots, y_n, z_n, \zb_n). \ee
Discrete labels such as particle type, helicity, and in/out are suppressed. It is related to the stripped amplitude $A_n$ as
\be \ca_n = \delta^{(4)} \Big( \sum_{i=1}^n p_i \Big) A_n. \ee
An $n$-point amplitude is na\"ively described by $3n$ continuous variables, but the momentum conserving delta function restricts us to a codimension-four locus. We can attempt to use $3n-4$ of our variables as coordinates on this locus, and generically this will be well-behaved. Furthermore most, though not all, choices of elimination will involve rational expressions of the other $3n-4$ variables. One choice giving rational substitutions is eliminating $\omega_i, \zb_i, \omega_j, \zb_j$ for some $i, j$, which is essentially the same as the standard spinor-helicity approach of eliminating two square bracket spinors \cite{Cachazo:2014fwa}. Other rational choices include eliminating $\omega_i, \zb_i, y_j, \zb_j$, or $y_i, \zb_i, y_j, \zb_j$, or $\omega_i, \omega_j, \omega_k, \omega_\ell$. For simplicity assume we make some such rational choice, and denote our $3n-4$ coordinates by $\xi^I$. These coordinates will not describe the locus globally, but they will be valid almost everywhere, which is sufficient for our purposes. The amplitude $A_n(\xi^I)$ will be a rational function whose only poles come from internal propagators going on shell, and places where the coordinate system breaks down, which we avoid. We now move on to a discussion of coordinate independence.

Suppose we want to compute the residue ${\rm Res}_{z_1\shortrightarrow z_*} \, \ca_n$ of the unstripped amplitude, where $z_*$ is any function of the $\xi^I$.\footnote{Assume $z_*$ can be approached such that the coordinates remain valid.} There are no double or higher poles, so let us define
\be \label{eq:resdef} \res{z_1\shortrightarrow z_*} \ca_n \equiv \big[ (z_1-z_*) \ca_n \big] \big|_{z_1=z_*}. \ee
The right hand side (RHS) makes no reference to the choice of coordinates $\xi^I$. Even if $z_1$ is eliminated, the RHS is still well-defined. Furthermore it is easy to see that if $z_1$ is one of our coordinates and $z_*$ is independent of $z_1$, then the RHS is equivalent to a conventional residue on the stripped amplitude $A_n$,
\be \big[ (z_1-z_*) \ca_n \big] \big|_{z_1=z_*} = \delta^{(4)} \Big( \sum_{i=1}^n p_i \Big) \res{z_1\shortrightarrow z_*} A_n. \ee
This is the version used in the body of this paper, but the equivalence to the RHS of \eqref{eq:resdef} guarantees that the result is independent of the choice of coordinates.

An insertion $\oh(\omega, z, \zb)$ is a function of $\omega, z, \zb$, along with the other $3n-7$ implicit coordinates. If we change our choice of elimination then the new $\oh(\omega, z, \zb)$ will be related to the old one simply by substitution on the newly eliminated coordinates. The same is not true of $\oh^{(k)}(z, \zb)$. If we change coordinates and substitute for the newly eliminated implicit coordinates, they will generically acquire $\omega$ dependence which must be expanded. In this way $\oh^{(k)}(z, \zb)$ is sensitive not only to its own implicit substitutions, but also those of $\oh^{(\ell)}(z, \zb)$ with $\ell>k$ (recall $\oh^{(k)}$ comes from the $\omega^{-k}$ term). It is very satisfying to see how this behavior is consistent with the soft theorems. In practice we will simply fix a choice of coordinates and not have to worry about these subtleties.

\section{Summary of commutativity of limits} \label{app:comm}

In this appendix we summarize some useful relations between Mellin transforms, soft limits, and residues in $z_{ij}$. As noted in section \ref{sec:ssoft}, soft limits are equivariant with the Mellin transform in the sense that taking an energetically soft limit on $\omega_i$ is equivalent to Mellin transforming on $\omega_i$ and then taking a conformally soft limit on $\Delta_i$. This means that consecutive energetically soft limits are equivalent to consecutive conformally soft limits. As discussed in section \ref{sec:hard}, the residue ${\rm Res}_{z_i\shortrightarrow z_j}$ commutes with all Mellin transforms and with soft limits on $\omega_i$ and $\omega_j$ (but not other $\omega_k$). Soft limits on $\omega_i, \omega_j$ do not commute in general.

The Mellin transform is simply a change of basis taking us from an unstripped momentum space amplitude to a celestial amplitude. In fact, there is no need to use the same basis for each leg. We are free to work with mixed amplitudes where some legs are momentum eigenstates and some are conformal primaries, although then the dependence on the momentum space insertions may no longer be rational.

\bibliography{jacobi}
\bibliographystyle{utphys}

\end{document}